\documentclass{PoS}

\usepackage{graphicx}
\usepackage{amsmath}
\usepackage{amssymb}
\usepackage{multirow}
\usepackage{pifont}

\DeclareMathOperator{\Tr}{Tr}

\title{Composite (Goldstone) Higgs Dynamics on the Lattice:
Spectrum of SU(2) Gauge Theory with two Fundamental Fermions}

\ShortTitle{Composite (Goldstone) Higgs Dynamics on the Lattice}

\author{Rudy Arthur\\
        E-mail: \email{rudy.d.arthur@gmail.com}}

\author{Vincent Drach\\
        E-mail: \email{drach@cp3-origins.net}}

\author{Martin Hansen\\
        E-mail: \email{hansen@cp3-origins.net}}

\author{\speaker{Ari Hietanen}\\
        E-mail: \email{hietanen@cp3-origins.net}}

\author{Randy Lewis\\
        E-mail: \email{randy.lewis@yorku.ca}}

\author{Claudio Pica\\
        E-mail: \email{pica@cp3-origins.net}}

\author{Francesco Sannino\\
        E-mail: \email{sannino@cp3-origins.net}}

\abstract{We study the meson spectrum of the SU(2) gauge theory with two
Wilson fermions in the fundamental representation. The theory unifies
both Technicolor and composite Goldstone Boson Higgs models of
electroweak symmetry breaking. We have calculated the masses of the
lightest spin one vector and axial vector mesons. In addition, we have
also obtained preliminary results for the mass of the lightest scalar 
(singlet) meson state. The simulations have been done with multiple
masses and two different lattice spacings for chiral and continuum
extrapolations. The spin one meson masses set lower limits for 
accelerator experiments, whereas the scalar meson will mix with a pGB of
the theory and produce two scalar states. The lighter of the states is
the 125 GeV Higgs boson, and the heavier would be a new yet unobserved
scalar state. \\\\
CP3-Origins-2014-046 DNRF90 \& DIAS-2014-46
} 

\FullConference{The 32nd International Symposium on Lattice Field Theory,\\
                23-28 June, 2014\\
                Columbia University New York, NY}

\begin{document}

\section{Introduction}
The SU(2) gauge theory with two fundamental Dirac fermions is the
simplest field theoretical realization of unified theory of Composite
Goldstone Boson Higgs (CGBH) and Technicolor \cite{Cacciapaglia:2014uja}. The
theory has a chiral symmetry breaking pattern from
SU(4)$\rightarrow$SP(4)$\sim$SO(5) which gives five Goldstone
Bosons \cite{Appelquist:1999dq,Ryttov:2008xe}. The breaking direction with respect to standard model can be
parametrized by an angle $\theta$. There are two extreme cases
$\theta=0$ and $\theta=\pi/2$. The former corresponds to a composite
Goldstone Boson Higgs model and the latter to a Technicolor model.

In the CGBH limit the electroweak symmetry is intact, i.e., it
resembles the standard model with a zero Higgs mass. The four components of the
Higgs doublet are composed of four Goldstone Bosons. The fifth GB is standard
model neutral. 

In the Technicolor limit the electroweak symmetry is completely
broken, with three of the five Goldstones eaten by the SM gauge
bosons. The Higgs particle is the lightest scalar $\sigma$ in the SU(2) gauge 
theory. The remaining GB form a complex doublet which is a possible
asymmetric dark matter candidate \cite{Ryttov:2008xe}.

The value of $\theta$ is determined dynamically. Besides the
electroweak and top induced radiative contributions its final value
depends also on new possible operators breaking the original SU(4)
global symmetry explicitly.  A generic value of $\theta$ between the
two extreme cases is therefore naturally expected \cite{Cacciapaglia:2014uja}.  
 In this case the composite GB Higgs state mixes with the Technicolor
$\sigma$ state giving two scalar particles of which the lightest
one is identified as the Higgs Boson. The scalar masses obtain
large corrections from the coupling to top
\cite{Foadi:2012bb,Cacciapaglia:2014uja} whereas the (axial)vector
meson masses are mostly determined by the dynamics of the SU(2) gauge
theory. The scale of the theory is $f_\Pi \sin \theta=246$ GeV.

Here we will review lattice calculations of the model performed in
\cite{Hietanen:2014xca} (see also \cite{Lewis:2011zb}) and present
preliminary results of measuring the mass of the scalar
meson. Scattering lengths of the same model has been studied in
\cite{Arthur:2014zda}.  In addition to the unified composite Goldstone
Boson dynamics the SU(2) theory with two fundamental fermions has been
used widely in beyond standard model phenomenology and dark matter
studies \cite{Ryttov:2008xe,Hietanen:2013fya,Detmold:2014kba,Buckley:2012ky}. 

\section{Theory}
We use the standard Wilson plaquette action with Wilson fermions.
\begin{eqnarray}
S_W &=& \frac{\beta}{2}\sum_{x,\mu,\nu}\left(1-\frac{1}{2}{\rm ReTr}U_\mu(x)
        U_\nu(x+\hat\mu)U_\mu^\dagger(x+\hat\nu)U_\nu^\dagger(x)\right)
      + \sum_x\overline{\psi}(x)(4+m_0)\psi(x) \nonumber \\
   && - \frac{1}{2}\sum_{x,\mu}\left(\overline{\psi}(x)(1-\gamma_\mu)U_\mu(x)\psi(x+\hat\mu)
   +\overline{\psi}(x+\hat\mu)(1+\gamma_\mu)U_\mu^\dagger(x)\psi(x)\right) \,, 
\end{eqnarray}
where $U_\mu$ is the gauge field and $\beta$ the gauge coupling in conventional
lattice notation.  $\psi$ is the doublet of $u$ and $d$ fermions, and
$m_0$ is the mass matrix.

We extract the non singlet meson masses from the two-point correlation functions
\begin{align}
C^{(\Gamma)}_{\overline{u}d}(t_i-t_f)
 & =  \sum_{\vec x_i,\vec x_f} \left\langle {\cal O}_{ud}^{(\Gamma)}(x_f)
{\cal O}_{ud}^{(\Gamma)\dagger}(x_i) \right\rangle\nonumber\\
 & = \sum_{\vec x_i,\vec x_f} \Tr \Gamma
S_{d\overline{d}}(x_f,x_i)\gamma^0\Gamma^\dagger\gamma^0 S_{u\overline{u}}(x_i,x_f),
\end{align}
where $S_{u\overline{u}}(x,y) = \langle
u(x)\overline{u}(y)\rangle$. The quantities of interest are
pseudoscalar $\Gamma=\gamma_5$, vector $\Gamma=\gamma_k$ ($k=1,2,3$),
and axial vector $\Gamma=\gamma_5\gamma_k$ mesons. We use stochastic
estimator for the non singlet correlators with $Z_2\times Z_2$ single
time slice stochastic sources. 

The correlator for  scalar meson reads
\begin{equation}
C_{\rm 2pts}(t)=\sum_{\vec{x}} \Tr\big\{ S(x,0) S(0,x)\big\} -
2\sum_{\vec{x}} \Tr\big\{S(x,x)\big\} \Tr\big\{S(0,0)\big\}.
\end{equation}
It obtains contributions from connected and disconnected part. 
To approximate the disconnected part, we use color
spin and volume diluted unit sources. Consider a source of type 
\begin{equation}
  \eta_{ac}(x) = \delta_{ac} \sum_{y}\delta(x,y),
\end{equation}
where $a,c$ labels both color and spin. Solving the Dirac equation
with this source gives
\begin{equation}
 \psi^c_a(x) =  \sum_{y} D^{-1}_{ab} (x,y)\eta_b^c(y) = \sum_y S_{ac}(x,y).
\end{equation}
Now if we multiply by the source and sum over $c$ and spatial volume
$\vec x$ we get
\begin{align}
  \sum_{\vec{x}} \eta_a^c({\vec x},t)\psi_a^c({\vec x},t) &= \sum_{{\vec x},y}\Tr[S(x,y)]
   = \sum_{\vec x} \Tr[S(x,x)],
\end{align}
which is the disconnected part we want to evaluate. The last equality
follows from Elitzur's theorem stating that the gauge average of a
gauge non-invariant object vanishes. 

The largest gauge noise from this source type arises from terms
$S(x,y)$, where $x$ and $y$ are nearest neighbors. This can be reduced
if we consider volume diluted source of type
\begin{equation}
\eta^c_{a,k,n}(x) =\delta_{ac}\sum_{y\in P(k,n)} \delta(x,y),
\end{equation}
where $P_{k,n} = \{x=(x_0,x_1,x_2,x_3)|x_0+x_1+x_2+x_3=k\mod n\}$,
i.e. the set $P_{0,2}$ is the set of even lattice sites and $P_{1,2}$ the odd
ones. Now again we have 
\begin{equation}
  \sum_{k=0}^{n-1} \sum_{x\in P_{k,n}}
  \eta^c_{a,k,n}(x)\psi^c_{a,k,n}(x) = \sum_{\vec x} \Tr[S(x,x)].
\end{equation}
However, the leading gauge non-invariant terms are of a distance $n$
apart and hence suppressed. Our lattice sizes $L=32\times16^2$ limits
us to $n=8$. Naturally we need to perform also $n$ times more
inversions.

%\begin{figure}
%  \begin{center}
%  \includegraphics[width=0.8\textwidth]{finite_vol_b2_2m-0_75.pdf}
%  \caption{The finite volume dependence of the
%    observables.\label{fig:fvol}}
%  \end{center}
%\end{figure}

\begin{figure}
  \begin{center}
  \includegraphics[width=0.8\textwidth]{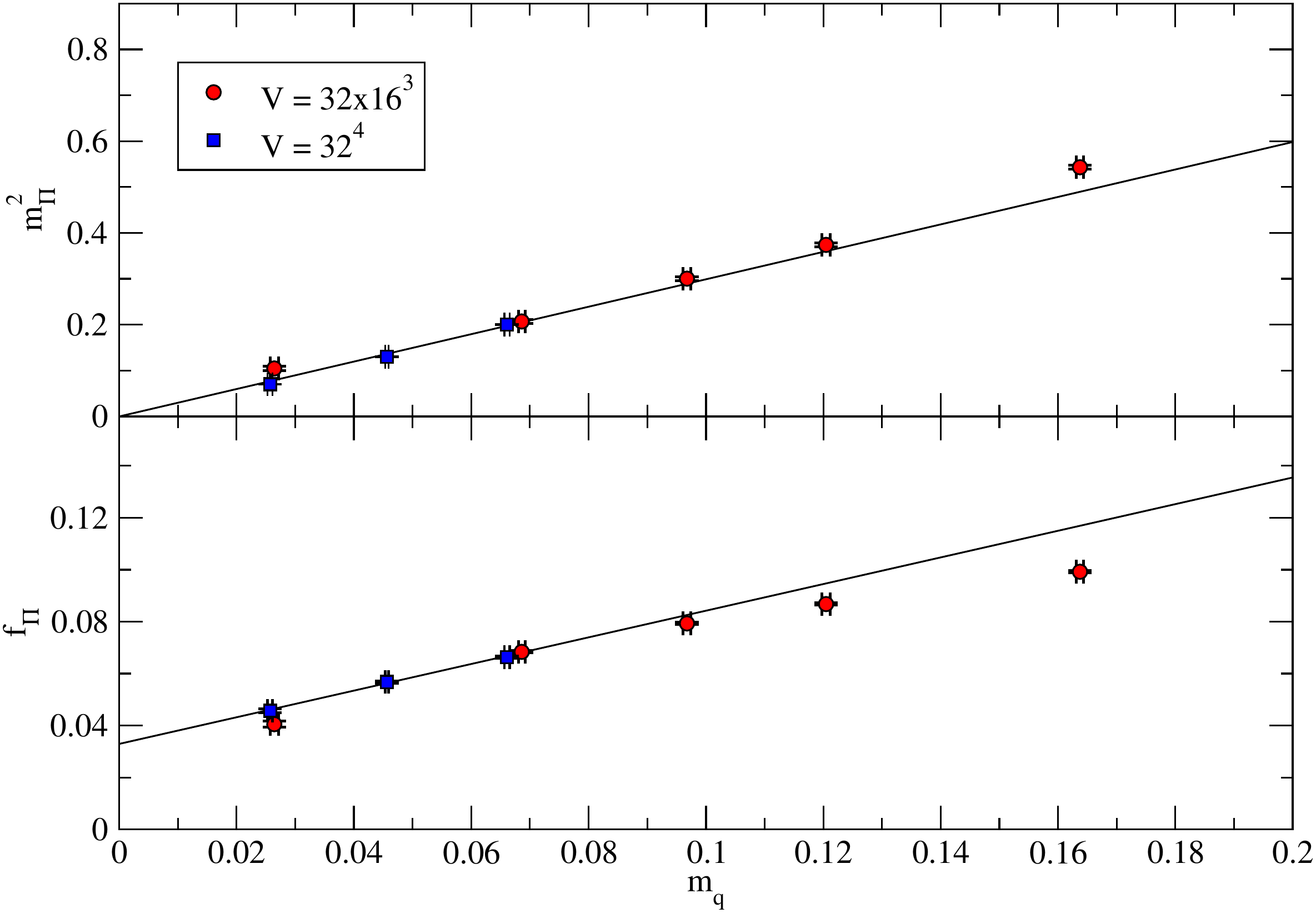}
  \caption{The chiral extrapolations of $m_\Pi$ and $f_\Pi$ as a
    function of the PCAC-mass $m_q$ in lattice units. \label{fig:chiral}}
  \end{center}
\end{figure}

\begin{figure}
  \begin{center}
  \includegraphics[width=0.445\textwidth]{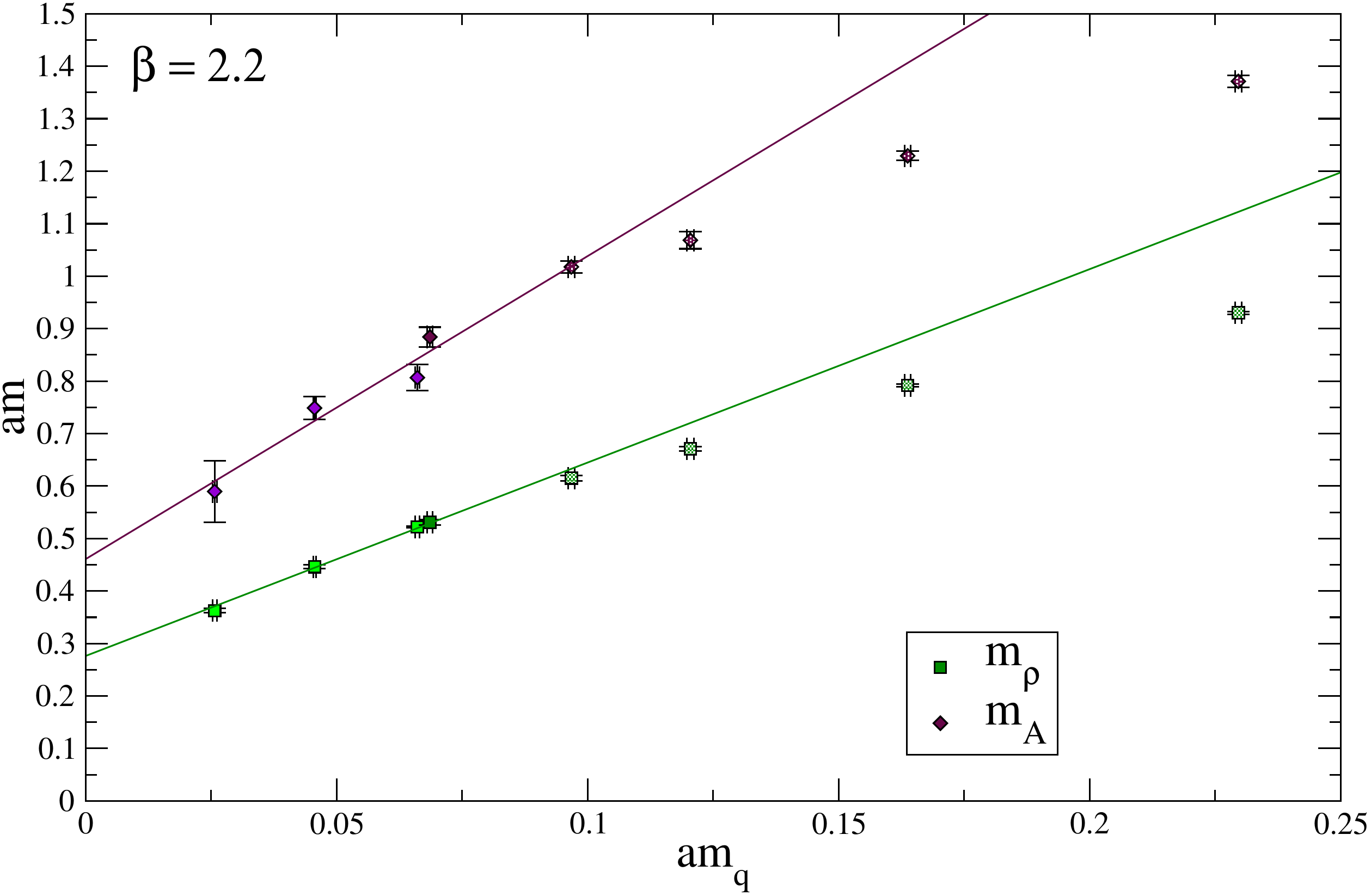}~~
  \includegraphics[width=0.41\textwidth]{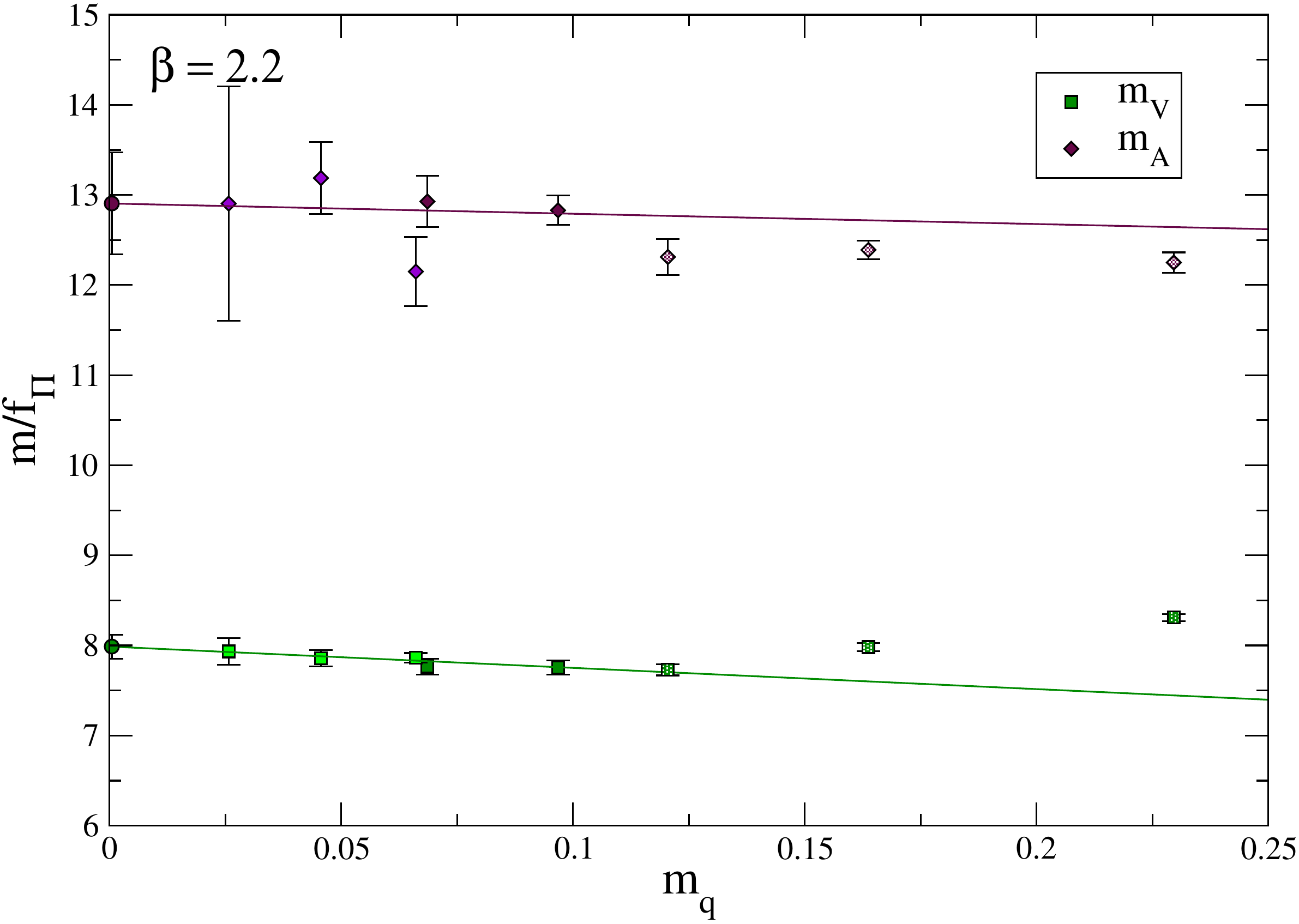}
  \caption{Left panel: Mass of vector and axial vector meson in lattice units for
    the finer lattice as a function of the PCAC-mass. Right panel:
    Mass of vector and axial vector meson as units of $f_\Pi$ for
    the finer lattice as a function of the PCAC-mass. \label{fig:mesons}}
  \end{center}
\end{figure}

\begin{figure}
  \begin{center}
  \includegraphics[width=0.8\textwidth]{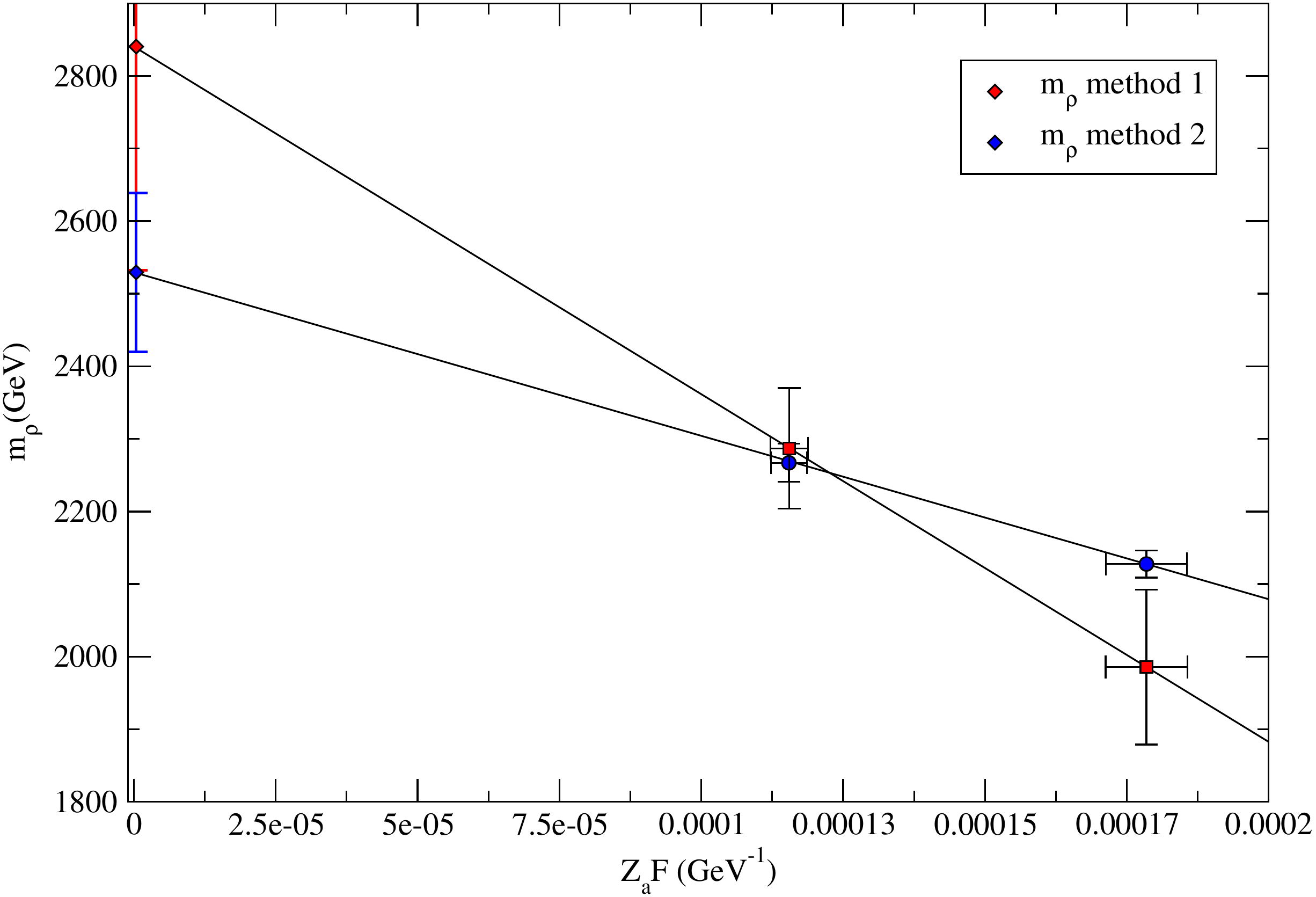}
  \caption{The continuum extrapolation of vector and axial vector
    meson masses as a function of lattice spacing.\label{fig:continuum}}
  \end{center}
\end{figure}

\begin{figure}
  \begin{center}
  \includegraphics[width=0.9\textwidth]{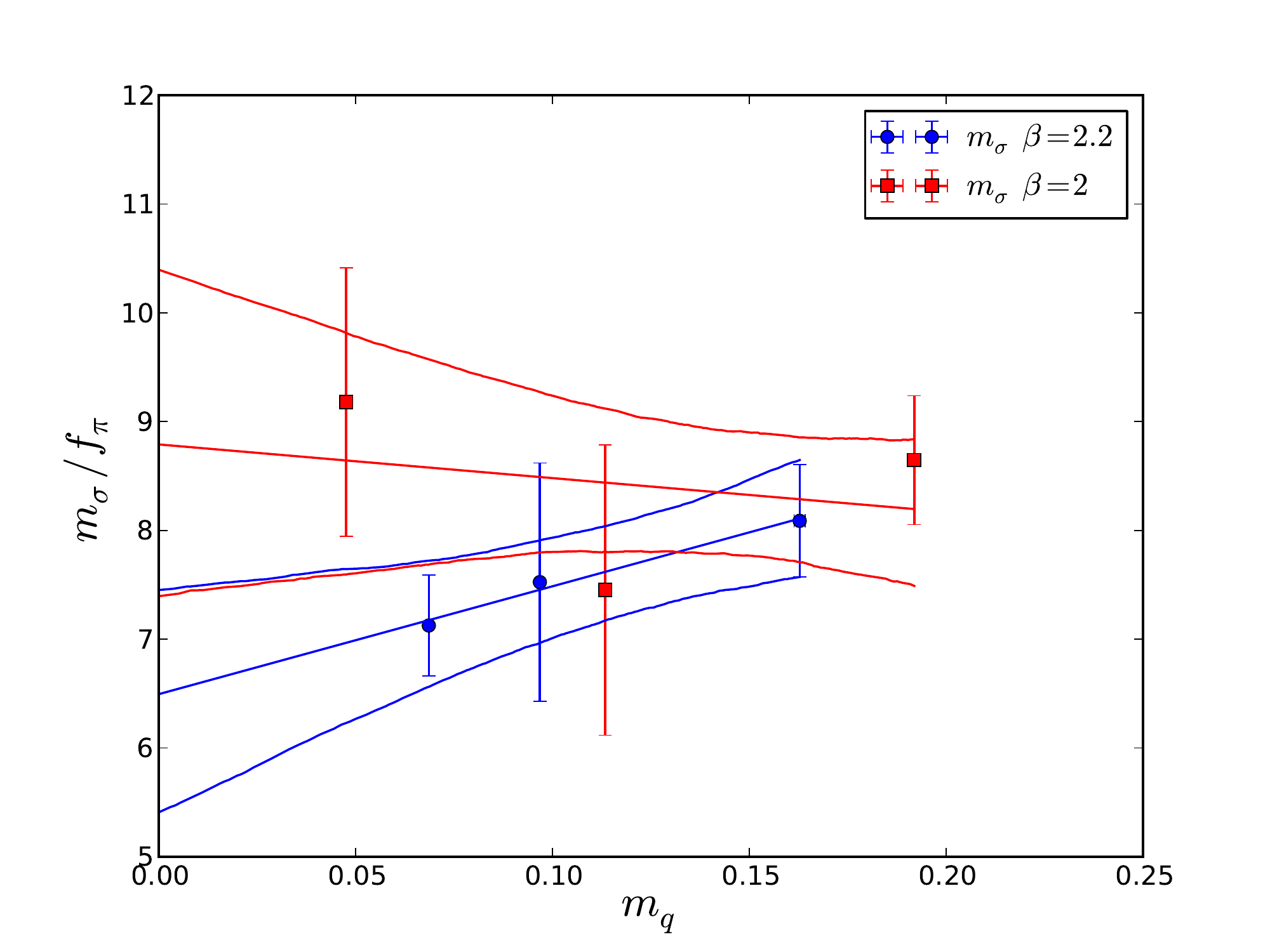}
  \caption{Scalar mass in the units of $f_\Pi$ as a function of the
    unrenormalized PCAC-mass $m_q$.\label{fig:scalar}}
  \end{center}
\end{figure}

\section{Results}
We have performed the simulations with two different lattice spacings
$\beta=2.0$ and 2.2, and with six and seven different fermion masses
respectively. We also performed some simulations varying volume to
control finite volume effects. In our most chiral point in finer
lattice we see finite volume effects on the smallest volume, but
on larger volumes they are smaller than our statistical error.

The chiral extrapolation works qualitatively well for $m_\Pi$ and
$f_\Pi$. See Fig.~\ref{fig:chiral}. However, fixing the known low-energy constants $C$
and $C'$ in the chiral perturbation theory result \cite{Bijnens:2009qm}
\begin{equation}
  \frac{m^2_{\rm \Pi}}{m_{\rm q}} = 2B\left[1 + C x\log x + D x +
  \mathcal{O}(m_{\rm q}^2)\right]\,\,,
  \label{eq:chiralgb}
\end{equation}
and 
\begin{equation}
  f_{\rm \Pi} = F\left[1 + C' x\log x + D' x +
  \mathcal{O}(m_{\rm q}^2)\right]\,\,,
  \label{eq:chiralfpi}
\end{equation}
where $B$, $F$, $D$ and $D'$ are unknown, and $x\equiv\frac{2B m_{\rm
    q}}{16\pi^2 F^2}$., does not describe the data well. 

The scale of the theory is set by requiring that $f_\Pi\sin\theta=246$
GeV. In the extrapolation to chiral limit the scale can be set in two 
different ways. We can extrapolate separately $m_{\rho,A}$ and $f_\Pi$
to the continuum limit and take the ratio afterwards (Fig.~\ref{fig:mesons}), or we can take 
ratio of $m_{\rho,A}$ and $f_\Pi$ first and then extrapolate to the
chiral limit (Fig.~\ref{fig:mesons}). The latter works better as the chiral extrapolation is
almost flat, hence we use it for the central value. However, they both
result to a same number within the statistical errors.

As we have only two different lattice spacings we can only perform a
linear fit to the continuum extrapolated data with little control over
systematic errors (see Fig.~\ref{fig:continuum}). Hence, we take a conservative approach, and
approximate the systematic error to be the difference between the
value on the finest lattice spacing and the continuum limit. Another
source of systematic error is an unknown renormalization factor $Z_a$,
which multiplies the $f_{\rm \Pi}$. As an estimate we use its
perturbative value $Z_a=1-0.2983/\beta$ \cite{DelDebbio:2008wb} and as an estimate of
systematic error the difference of results between perturbative $Z_a$
and using $Z_a=1$. This leads to our final results:
\begin{align}
  m_\rho\sin\theta&= 2520(100)(240)(310)\approx2500(500)GeV\nonumber\\
  m_A\sin\theta&=3300(400)(510)(340)\approx3300(700)GeV\nonumber.
\end{align}

For a reliable measurement of the scalar meson mass we need an order
of 10 000 configurations, and we are still in process of generating
them. The connected part can be identified with isovector 
scalar meson and our data shows that it is heavier than the
disconnected part and can be neglected. Our preliminary results are
shown in Fig.~\ref{fig:scalar} with two different lattice spacings and
suggest that the scalar is lighter than the vector meson. Analytical
 predictions for the scalar mass, before electroweak and top
 corrections, appeared in \cite{Foadi:2012bb} using large N scaling
 arguments and it is expected to be approximately around 1 to 1.5
 TeV. These simple estimates are also supported by very recent results
 \cite{Vujinovic:2014ioa} making use of Dyson-Schwinger
 approximations.    

\section{Summary}
 We have provided phenomenological relevant information for the
 spectrum of an underlying gauge theory naturally unifying at the more
 fundamental level models of CGBH and Technicolor. The continuum
 extrapolated vector meson mass is $m_\rho\sin\theta =2.5(5)$TeV. The largest errors come
 from the continuum extrapolation and unknown $Z_a$, we are in a
 process of improving this calculation. The preliminary results
 indicate that the scalar meson would be lighter than the
 $\rho$-meson. However, more statistics is needed to confirm this
 observation.  Our preliminary results also hints that the lightest
 new state of the theory could be another scalar particle.

\acknowledgments

This work was supported by the Danish National Research Foundation
DNRF:90 grant, by a Lundbeck Foundation Fellowship grant, and NSERC of
Canada. The computing facilities were provided by the Danish Centre
for Scientific Computing.

\end{document}